\documentclass[aps,reprint,superscriptaddress,floatfix,longbibliography]{revtex4-2}

\usepackage{slashed}
\usepackage{amsmath}
\usepackage{mathtools}
\usepackage{amssymb}
\usepackage{graphicx,bm}
\usepackage{color}
\usepackage{hyperref}
\usepackage[squaren]{SIunits}
\usepackage{microtype}

\usepackage{ulem}

\usepackage[dvipsnames]{xcolor}

\usepackage{verbatim}

\usepackage{grffile}

\graphicspath{{./figures/}}

\newcommand{\Eqref}[1]{Eq.~\eqref{#1}}
\newcommand{\Lagr}{\mathcal{L}}

\usepackage{hyperref}
\hypersetup{
    colorlinks=true,
    linkcolor=blue,
    citecolor=red,
    filecolor=magenta,      
    urlcolor=cyan,
}

\begin{document}

\title{Numerical optimization of quantum vacuum signals}

\author{Maksim Valialshchikov}
\email[]{maksim.valialshchikov@uni-jena.de}
\affiliation{Helmholtz Institute Jena, Fröbelstieg 3, 07743 Jena, Germany}
\affiliation{Institute of Optics and Quantum Electronics, Friedrich-Schiller-Universität, Max-Wien-Platz 1, 07743 Jena, Germany}

\author{Felix Karbstein}
\email[]{felix.karbstein@uni-jena.de}
\affiliation{Helmholtz Institute Jena, Fröbelstieg 3, 07743 Jena, Germany}
\affiliation{GSI Helmholtzzentrum für Schwerionenforschung GmbH, Planckstrasse 1, 64291 Darmstadt, Germany}
\affiliation{Theoretisch-Physikalisches Institut, Abbe Center of Photonics, Friedrich-Schiller-Universit\"at Jena, Max-Wien-Platz 1, 07743 Jena, Germany}

\author{Daniel Seipt}
\email[]{d.seipt@hi-jena.gsi.de}
\affiliation{Helmholtz Institute Jena, Fröbelstieg 3, 07743 Jena, Germany}
\affiliation{GSI Helmholtzzentrum für Schwerionenforschung GmbH, Planckstrasse 1, 64291 Darmstadt, Germany}
\affiliation{Institute of Optics and Quantum Electronics, Friedrich-Schiller-Universität, Max-Wien-Platz 1, 07743 Jena, Germany}

\author{Matt Zepf}
\affiliation{Helmholtz Institute Jena, Fröbelstieg 3, 07743 Jena, Germany}
\affiliation{GSI Helmholtzzentrum für Schwerionenforschung GmbH, Planckstrasse 1, 64291 Darmstadt, Germany}
\affiliation{Institute of Optics and Quantum Electronics, Friedrich-Schiller-Universität, Max-Wien-Platz 1, 07743 Jena, Germany}

\date{\today}

\begin{abstract}
    
The identification of prospective scenarios for observing quantum vacuum signals in high-intensity laser experiments requires both accurate theoretical predictions and the exploration of high-dimensional parameter spaces. Numerical simulations address the first requirement, while optimization provides an efficient solution for the second one.
In the present work, we demonstrate the potential of Bayesian optimization in maximizing photonic quantum vacuum signals on the example of two-beam collisions. This allows us to find the optimal waist sizes for beams with elliptic cross sections, and to identify the specific physical process leading to a discernible signal in a coherent harmonic focusing configuration scenario.
    
\end{abstract}

\maketitle

\section{Introduction}

Quantum vacuum fluctuations mediate effective nonlinear interactions between macroscopic electromagnetic fields and impact the dynamics of the latter. Within the Standard Model of particle physics, the leading quantum vacuum nonlinearities are governed by quantum electrodynamics (QED) and arise from the coupling of electromagnetic fields by a virtual electron-positron loop \cite{fedotov2023advances,marklund2006nonlinear,king2016measuring,karbstein2020probing}. 
The microscopic origin of these effects is a fundamental quantum process known as ``photon-photon scattering'' or ``light-by-light scattering'' \cite{Karplus:1950zza,Karplus:1950zz, DeTollis:1964una} that naturally affects the polarization, wavevector and frequency of the incident photons.

Modern laser and detector technologies allow to propose \cite{King:NatPhot2010,king2012photon,shen2018exploring,karbstein2018vacuum,wang_exploring_2024} and perform \cite{Moulin:1996vv,Moulin:1999hwj,Bernard:2000ovj} first photon-photon scattering experiments; see also the reviews \cite{marklund2006nonlinear,DiPiazza:RevModPhys2012,king2016measuring,karbstein2020probing,fedotov2023advances, Battesti:2012hf}. 
Theoretical studies mainly based on analytical considerations improved our understanding of the dependence of the quantum vacuum signals on the collision geometry and  the parameters of the incident laser fields.
However, the possibility of such insights typically relies on simplifying assumptions about the laser beam profile, like e.g. the use of an {\it infinite Rayleigh range approximation} \cite{Gies:2017ygp,King:2018wtn}.

One possibility to go beyond this known territory and to accurately predict quantum vacuum signals in the field configurations actually realized in experiment is the use of numerical Maxwell solvers. In \cite{lindner2023numerical} the authors propose a classical nonlinear Maxwell solver which could scale to arbitrary accuracy and include higher order nonlinear interactions of the field but requires a lot of computational resources. The authors of \cite{grismayer2021quantum} present a nonlinear Maxwell solver based on a generalized Yee scheme.
Their approach is less computationally expensive than \cite{lindner2023numerical} but includes interpolation of fields in the spatio-temporal domain and uses a lower accuracy order.
In the present work, we resort to the numerical approach put forward by \cite{blinne2019all}, that combines the {\it vacuum emission picture} \cite{karbstein2015stimulated} with a numerical solver for the linear Maxwell equations governing the dynamics of the initially applied classical electromagnetic fields; see also also \cite{Sainte-Marie:2022efx} for an alternative implementation.

The analysis of multiple new collision configurations, or the thorough study of the experimentally most relevant ones, usually requires the exploration of large parameter spaces to identify the optimal parameters that maximize a given observable. Especially the formidable computational costs of direct numerical simulations on a grid motivate the use of optimization methods that can efficiently guide the search.
Additionally, for currently available technologies, expected light-by-light scattering signal is typically weak (only a few photons; this also justifies the neglect of backreaction) against a huge laser background which makes it even more crucial to identify optimal configurations.

In this article, we apply a modern optimization framework to the nonlinear QED signature of photon-photon scattering in all-optical laser beam collisions (Section \ref{sec:formalism}). 
We perform numerical simulations (Section \ref{sec:numerics:simulation}) of several collision scenarios where Bayesian optimization guides the parameter search (Section \ref{sec:numerics:bayesian}). 
Resorting to known results previously studied in the literature, we benchmark and demonstrate the reliability of optimization. We in particular show that our approach allows us to more accurately determine the optimal parameters for the all-optical two-beam collision scenario studied in \cite{gies2022all} (Section \ref{sec:elliptic}). Moreover, we detail how it allows us to identify and resolve the origin of a promising discernible signal in a coherent harmonic focusing configuration \cite{karbstein2019boosting} (Section \ref{sec:harmonics}).

Aside from a few exceptions where we intend to highlight the explicit dependence on $c$ and $\hbar$, throughout this work, we use the Heaviside-Lorentz system with natural units $\hbar=c=\epsilon_0=1$.

\section{Formalism}
\label{sec:formalism}
The Heisenberg-Euler Lagrangian ${\cal L}_{\rm HE}$ \cite{Heisenberg:1936nmg} encodes quantum corrections to Maxwell's classical theory of electromagnetic fields in vacuo in effective nonlinear interactions of the applied electromagnetic fields.
Its leading nontrivial contribution in the small-field and low-frequency limit reads \cite{euler1935scattering}
\begin{equation}
    \Lagr_{\text{HE}}^{1\text{-loop}} \simeq \frac{m_e^4}{360 \pi^2} \left(\frac{e}{m_e^2}\right)^4 (4 \mathcal{F}^2 + 7 \mathcal{G}^2)\,,
    \label{eq:HE-lagrangian}
\end{equation}
where $e$ is the elementary charge, $m_e$ is the electron mass, and
$\mathcal{F}=\frac{1}{4} F_{\mu \nu} F^{\mu \nu} = \frac{1}{2} (\mathbf{B}^2 - \mathbf{E}^2)$, $\mathcal{G}=\frac{1}{4} F_{\mu \nu} \prescript{\star}{ }{F^{\mu \nu}}=-(\mathbf{B} \cdot \mathbf{E})$ are the electromagnetic field invariants. 
To be precise, \Eqref{eq:HE-lagrangian} holds for fields that fulfill $\{|\mathbf{E}|,c|\mathbf{B}|\} \ll E_{\rm cr}$, with critical electric field strength $E_{\rm cr} = m_e^2 c^3 / (e \hbar) \simeq 1.3 \times 10^{18}\,{\rm V/m}$, and vary on typical spatiotemporal scales $\lambda\gg\lambdabar_{\rm C}$, with electron Compton wavelength $\lambdabar_{\rm C}=\hbar/(m_ec)\simeq3.8\times10^{-13}\,{\rm m}$. These constraints are well satisfied by the fields available in state-of-the-art and near-future all-optical high-intensity laser experiments, which we consider in this paper.

As detailed in \cite{karbstein2015stimulated}, the zero-to-single signal photon transition amplitude to a state characterized by a wavevector $k^\mu=(\omega,{\bf k})$, with $\omega=|{\bf k}|$, and a polarization vector $\epsilon_{(p)}^{\mu}(k)$ can be expressed as
\begin{equation}
    S_{(p)}(\mathbf{k}) = \frac{\epsilon_{(p)}^{*\mu}(k)}{\sqrt{2 k^0}} \int d^4 x\, e^{ik_{\nu} x^{\nu}} j_{\mu}(x)\bigg|_{k^0=\omega},
    \label{eq:amplitude}
\end{equation}
where
\begin{equation}
    j_{\mu}(x) = 2 \partial^{\nu} \frac{\partial \Lagr_{\text{HE}}}{\partial F^{\nu \mu}}
\end{equation}
is the signal-photon current induced by the applied macroscopic electromagnetic fields $F^{\mu \nu}$. 

The differential number of signal photons is given by modulus squared of Eq.~\eqref{eq:amplitude},
\begin{equation}
    d^3 N_{(p)}(\mathbf{k}) = \frac{d^3 \mathbf k}{(2\pi)^3} |S_{(p)}(\mathbf{k})|^2.
\end{equation}
For a polarization insensitive measurement we sum over the signal photon polarization states, $d^3 N(\mathbf{k}) = \sum_p d^3 N_{(p)}(\mathbf{k})$. In this work we examine only polarization unresolved signals.
 Upon integrating over all possible values of the signal-photon energy, the polarization insensitive angular resolved signal photon density can be cast in the compact form
\begin{equation}
    \frac{d^2 N}{d^2\Omega}(\vartheta, \varphi) = \int_0^\infty \frac{d\omega \, \omega^2}{(2\pi)^3} \: \sum_p |S_{(p)}(\mathbf{k})|^2\,,
\end{equation}
with solid angle element $d^2\Omega=d\cos\vartheta\,d\varphi$. 

We denote the analogous quantities for the photons constituting the driving laser fields by $d^3 N^{\rm bgr}({\bf k})$ and $d^2N^{\rm bgr}({\bf k})/d^2\Omega(\vartheta,\varphi)$, respectively. These parameterize the background (``bgr") against which the signals have to be discriminated. Regions in momentum space where 
\begin{equation}
    \frac{d^3 N(\mathbf{k})}{d^3\mathbf k} > \frac{d^3 N^{\text{bgr}}(\mathbf{k})}{d^3 \mathbf k}
\end{equation}
are called \textit{spectrally discernible} and regions where
\begin{equation}
    \frac{d^2 N}{d^2\Omega}(\vartheta, \varphi) > \frac{d^2 N^{\text{bgr}}}{d^2\Omega}(\vartheta, \varphi)
\end{equation}
are called \textit{angularly discernible}. Upon integrating over these discernible regions we obtain the total discernible signal -- $N_{\text{disc}}$.

Throughout this work we study the interaction of two near-infrared intense laser pulses colliding under an angle of $90^\circ<\vartheta_{\text{col}}<180^\circ$.
Our focus is on two specific scenarios that were put forward recently. These envision the use of 1) beams with \textit{elliptic cross section} \cite{karbstein2016probing,gies2022all} (see Section \ref{sec:elliptic}), and 2) \textit{coherent harmonic focusing} \cite{karbstein2019boosting} (see Section \ref{sec:harmonics}) to enhance the quantum vacuum signals accessible in experiment. In the present study, we choose the collision geometry and laser parameters similar to those employed in Refs.~\cite{gies2022all,karbstein2019boosting}.

\section{Numerical details}
\label{sec:numerics}
\subsection{Simulation}
\label{sec:numerics:simulation}
To calculate signal photon spectrum, we employ the numerical code ``VacEm'' presented in \cite{blinne2019all} which uses 1) a linear Maxwell solver to describe the evolution of the external electromagnetic fields and 2) the vacuum emission picture to 
determine the leading quantum vacuum signal via Eq.~\eqref{eq:amplitude}. Self-consistent Maxwell propagation of external fields allows to avoid simplifying assumptions and approximations concerning the space-time structure of the focused laser fields typically required in analytical studies. This enables qualitatively accurate results in experimentally realistic field configurations.

We initialize the self-consistent Maxwell propagation by defining a model field configuration at focus ($t=0$) either in space or frequency domain (for details see Section III D of \cite{blinne2019all}). This field is then propagated to other time steps according to the linear Maxwell equations.  

For the \textit{elliptic cross section} scenario we model the pulses at focus as leading order paraxial Gaussian beams. For \textit{coherent harmonic focusing} we instead use a spectral laser pulse model \cite{waters2017beam,blinne2019all} for the initial data. In the limit of weak focusing and long pulse duration the latter converges to the zero-order paraxial Gaussian beam result. We made these particular choices to enable a better comparison with the results of \cite{gies2022all, karbstein2019boosting}.

\subsection{Bayesian optimization} 
\label{sec:numerics:bayesian}
The core idea of Bayesian optimization is to efficiently sample the parameter landscape given a limited resource budget \cite{shahriari2015taking, frazier2018tutorial}. This approach is especially useful in cases where the \textit{target functions} (i.e., those to be optimized) are very costly to evaluate; it also works for \textit{black-box} functions when neither an analytic expression nor gradient information is available. 

In Bayesian optimization, a statistical model represents the \textit{black-box} target function $f(x)$. Each new observation updates the model via Bayes' theorem, thus, incorporating all known data about $f(x)$.
A \textit{utility function} complements the statistical model and judges the prospects of particular points in parameter space for future observations.
It regulates the trade-off between exploration (survey of unknown regions in parameter space) and exploitation (choice of likely local maxima) and suggests promising candidates for the next observation. 

\begin{figure*}[t]
    \centering
    \includegraphics[width=0.75\linewidth]{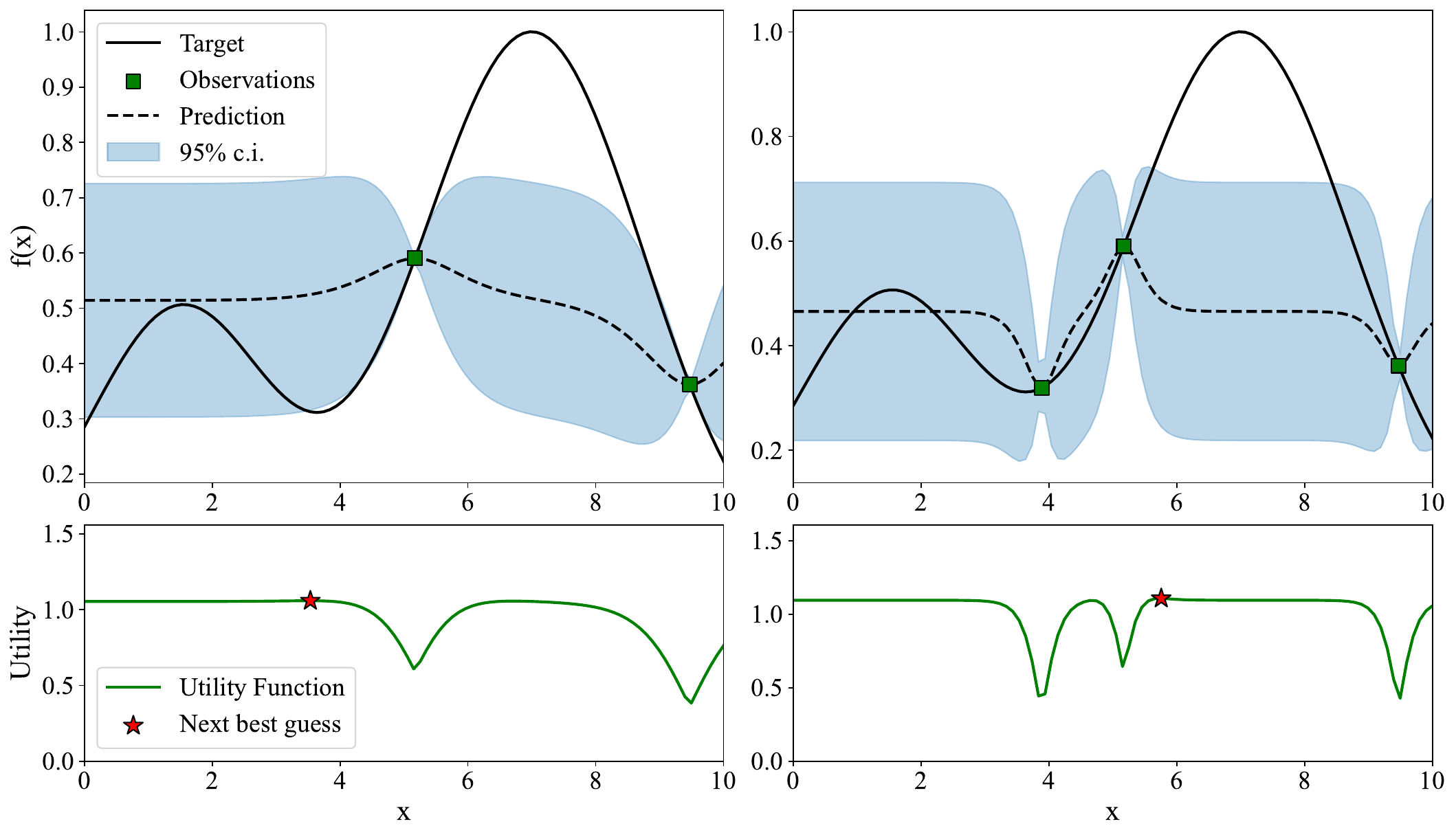}
    \caption{Schematic visualization of the Bayesian optimization process. (Left) For given initial observations, a Bayes model is initialized with certain mean and variance. The maximum of the \textit{utility function} determines the next observation. (Right) A new observation updates the model prediction and utility function. (Top) The black solid line illustrates the \textit{target function} and the green squares mark available observations. The dashed black line and blue area correspond to the mean and 95\% confidence interval of the Bayes model, respectively. (Bottom) \textit{Utility function} and estimated next best guess (candidate for future observation). Visualized with Python library \cite{bayesopt}.}
    \label{fig:bayes_opt}
\end{figure*}

Several steps of the Bayesian optimization procedure are shown in Figure \ref{fig:bayes_opt}. We start our search for the maximum of the target function by initializing a statistical model with constant mean and variance that serves as our initial prediction. Observations update this prediction by changing its mean and shrinking confidence intervals at their locations. They also update the \textit{utility function}, the maximum of which determines the next observation point. The update step is repeated after a new observation point is chosen. This procedure continues until convergence to the optimum is reached, or the computational budget is exhausted.

To explore promising light-by-light scattering scenarios, we use the optimization framework ``Optuna'' \cite{akiba2019optuna} which includes a high-level implementation of Bayesian optimization. In particular, it implements a tree-structured Parzen estimator approach which was designed for high-dimensional problems and modest computational budgets \cite{bergstra2011algorithms}. 

We start the optimization with 5-20 randomly sampled observations serving as a foundation for the Bayesian model initialization and continue up to several tens to few hundreds of Bayes steps. We parallelize the optimization procedure over several computational nodes. This is less efficient than the sequential approach in terms of computational resources usage, but allows to obtain more data in less time. 

\section{Circular and elliptic beam cross sections}
\label{sec:elliptic}

In this section both laser pulses are linearly polarized and characterized by a central oscillation frequency of $\omega_0 = 1.55\,$eV (wavelength $\lambda_0 = 800\,$nm), a duration ($1/{\rm e^2}$ wrt. intensity) of $\tau = 25\,$fs and an energy of $W = 25\,$J, unless specified otherwise.
We model the fields at focus ($t = 0$) in the spatial domain as leading order paraxial Gaussian from which we construct field spectral amplitudes that manifestly solve the linear Maxwell equations (for details see Section III D of \cite{blinne2019all}). Then the field is propagated to other time steps according to the exact linear Maxwell equations. The transverse profile of such pulses is characterized by two, in general different, waist sizes $w_{x} = \mu_x \lambda_0$ and $w_{y} = \mu_y \lambda_0$ in $x$ and $y$ direction, respectively; for circularly symmetric beams this reduces to $w=w_x=w_y = \mu \lambda_0$. In the diffraction limit where the minimal waist fulfilling $w \simeq \lambda_0$ is reached we have $\mu = 1$.

Without loss of generality, we assume the optical axis of each pulse to lie in the $xz$ plane. The associated wavevectors are given by $\mathbf{k_{\ell}} = \omega_0 \hat{\mathbf{k}}_{\ell}$, where $\hat{\mathbf{k}}_{\ell}$ with $\ell\in\{1,2\}$ are unit-vectors: $\hat{\mathbf{k}}_1 = \mathbf{e}_z$, $\hat{\mathbf{k}}_2 = \sin{\vartheta_{\text{col}}}\, \mathbf{e}_x + \cos{\vartheta_{\text{col}}}\, \mathbf{e}_z$. Laser pulse $\ell=1$ is polarized along the $x$-axis ($\mathbf{E}_1(t,\mathbf{x}) \:||\: \mathbf{e}_x, \mathbf{B}_1(t,\mathbf{x}) \:||\: \mathbf{e}_y$). The polarization of laser pulse $\ell=2$ is not fixed {\it a priori} and can be changed by tuning the angle $\beta_2=\sphericalangle \{\mathbf{E}_1(0,\mathbf{0}), \mathbf{E}_2(0,\mathbf{0})\}$ appropriately. The total polarization insensitive signal is maximized for $\beta_2 = 90^{\circ}$ which corresponds to a polarization along $\mathbf{e}_y$ for the $\ell=2$ pulse in a counter-propagating geometry, $\vartheta_{\rm col}=180^{\circ}$. In what follows we always assume $\beta_2 = 90^{\circ}$ unless mentioned otherwise.

The recent works \cite{karbstein2016probing, mosman2021vacuum,gies2022all} studied the influence of the probe waist size on the discernible signal in two-beam collision scenarios in the all-optical and x-ray optical regimes. They found that maximal focusing typically maximizes the total signal. On the other hand, weaker focusing tends to enhance the discernible signal. The latter is more relevant for experimental searches because it generically results in different far-field angular divergences of the background and signal fields; if the latter is wider, it provides a means to separate the signal from the background. Using an infinite Rayleigh range approximation to simplify the calculations, \cite{gies2022all} examines the dependence of the discernible signal and its directional emission characteristics on the collision angle and waist sizes of the colliding laser beams. For beams with circularly symmetric cross sections (1d parameter space), \cite{gies2022all} manages to identify the optimal waist size of the probe, while for beams with elliptic cross sections (2d parameter space) a hypothesis for the optimal waist size is invoked to avoid a full parameter scan. In this section we aim to compare our numerical results with the findings of \cite{gies2022all} and to refine the values for the optimal parameters extracted there with automatic optimization.

For this scenario the pump is focused to its diffraction limit unless mentioned otherwise. At the same time the probe waists in $x$ and $y$ directions are varied independently. In this section we are interested in optimizing \textit{angularly discernible} signals. The main advances of our approach compared to \cite{gies2022all} can be summarized as follows: 1) we use a self-consistent linear Maxwell solver to describe the propagation of the laser fields instead of resorting to a leading-order paraxial Gaussian beam model simplified further by invoking an infinite Rayleigh range approximation. 2) For pulses with elliptic cross sections we find the optimum in full parameter space rather than in a specific subspace constrained by {\it ad hoc} assumptions.

\begin{figure}
    \centering
    \includegraphics[width=\linewidth]{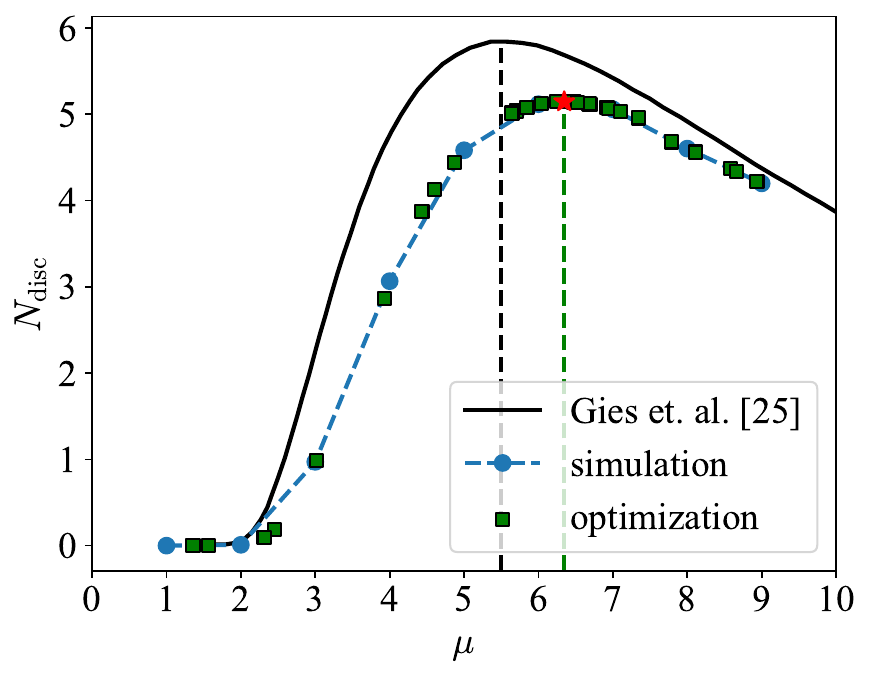}
    \caption{Number of discernible signal photons as a function of the probe waist size. The results presented here are for two beams (energy $W=25\,{\rm J}$, duration $\tau=25\,{\rm fs}$ each) of frequency $\omega_0$ colliding at an angle of $\vartheta_{\text{col}}=160^\circ$. The pump (probe) has a circular cross section with fixed (variable) value. The solid black line corresponds to analytical model calculations from \cite{gies2022all}, the dashed blue line to simulation results on a grid, and the green squares to simulation results from the optimization procedure. The red star highlights the optimal waist size found from optimization.}
    \label{fig:elliptic/w0_theta_160}
\end{figure}

One of the main findings of \cite{mosman2021vacuum,gies2022all} was that to maximize the total \textit{discernible} signal one needs to find a compromise between maximizing the intensity of the pulses in the interaction region and decreasing the background in the detection region. For beams with circular cross section this compromise results in an optimal waist size larger than its diffraction limit. This is illustrated in Figure \ref{fig:elliptic/w0_theta_160} which shows how changing the probe waist size affects the total discernible signal yield $N_{\text{disc}}$. The black line represents the result of an analytical calculation resorting to an infinite Rayleigh approximation taken from \cite{gies2022all}. It is surprising how closely it resembles simulation results obtained with a numerical Maxwell solver. Both the initial rise and later ``linear'' decay of $N_{\text{disc}}$ with $\mu$ are captured by the analytical calculation and the agreement with the outcome of the numerical simulation becomes better for weaker focusing. However, the results slightly differ by an overall scaling factor and in the location of the ``true'' optimal waist size. 

The considered optimization problem has a 1-dimensional parameter space, namely the waist size of the  circularly symmetric probe. Since the search space is straightforward, the optimum could be found directly from a grid scan, which serves us as a benchmark. Note that for higher dimensional parameter spaces grid scans might be unfeasible because individual simulations are computationally expensive. The data points marked by green squares in Fig. \ref{fig:elliptic/w0_theta_160} are obtained with the optimization procedure introduced above. The clustering of simulation data around the optimum originates in the fact that after an initial exploration of the full parameter space, the optimization process focuses on \textit{promising} regions having a higher \textit{expected}  value of the discernible signal. Because the optimization procedure is ``gridless'' for parameters of interest (it samples points from a continuous distribution) it allows to find the optimum with better accuracy for limited computational budget. 

The second half of \cite{gies2022all} examines probe beams with elliptic cross section. Tightly focusing the beam in one direction provides the required high intensity in the interaction volume with the pump, while the weaker focusing in perpendicular direction assures that the signal remains discernible. In this case the parameter space is 2-dimensional ($\mu_x, \mu_y$). However, for convenience and to minimize the computational cost \cite{gies2022all} replaced the 2d parameter scan with two consecutive 1d parameter scans. Of course, this usually does not lead to the location of global optimum.

\begin{figure}
    \centering
    \includegraphics[width=\linewidth]{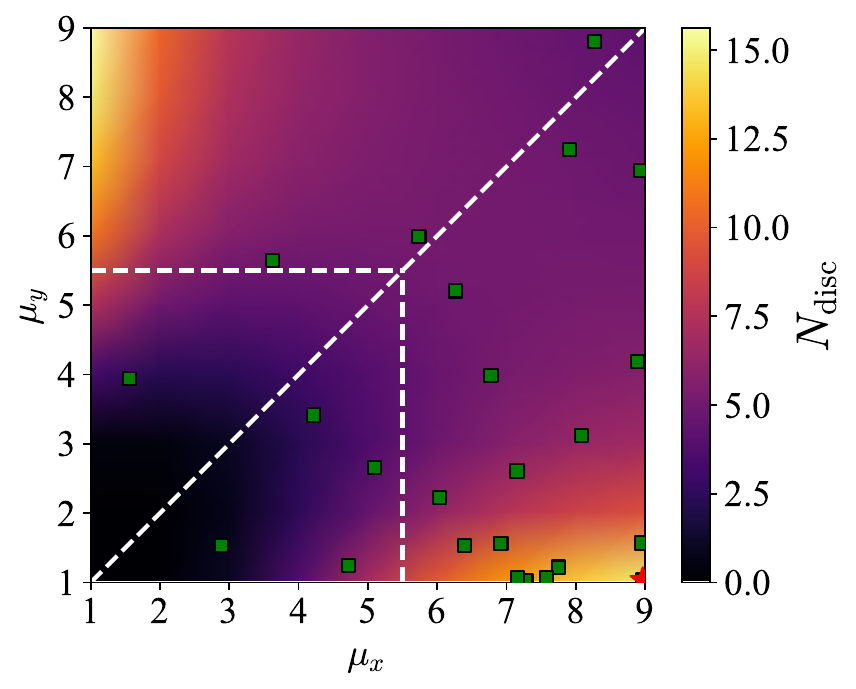}
    \caption{Number of discernible signal photons as a function of the two independent waist sizes for a probe with elliptic cross section. The results presented here are for two beams (energy $W=25\,{\rm J}$, duration $\tau=25\,{\rm fs}$ each) of frequency $\omega_0$ colliding at an angle of $\vartheta_{\text{col}}=160^\circ$. Pump (probe) has circular (elliptic) waist with constant (variable) value. The dashed white lines mark the one-dimensional parameter regions studied in \cite{gies2022all} (the simulation results depicted in Fig. \ref{fig:elliptic/w0_theta_160} are extracted from a diagonal slice of the colormap), green squares correspond to optimization trials, and the red star highlights the optimal waist size found from optimization.}
    \label{fig:elliptic/w0x_w0y_theta_160}
\end{figure}

Fig. \ref{fig:elliptic/w0x_w0y_theta_160} shows the dependence of the total discernible signal on the two independent probe waist sizes. It unveils two local maxima that are situated in the corners of the explored parameter space: one located at $(\mu_x=1, \mu_y=9)$ yielding $N_{\text{disc}} \approx 15.6$ and one at $(\mu_x=9, \mu_y=1)$ yielding $N_{\text{disc}} \approx 14.8$. The relative difference between these numbers is about $6\%$. Here, we constrained the explored parameter space for computational reasons. In turn, the ``true'' maxima might even lie beyond the chosen boundaries. This indeed turns out to be the case; see also the discussion in the context of Fig.~\ref{fig:theta_180_polarization} below. The dashed white lines mark the one-dimensional regions explored in \cite{gies2022all}. These clearly do not reach the real optimum. They suggested the optimal value ($\mu_x=5.5, \mu_y=1$) which gives $N_{\text{disc}} \approx 8.4$ while the maximal total discernible signal in the parameter space explored here is $N_{\text{disc}} \approx 15.6$. Note that this value is almost twice larger.  

As a reference, in Fig.~\ref{fig:elliptic/w0x_w0y_theta_160} we also provide simulation data (colormap) for the (appropriately discretized) whole considered parameter space, even though the computational cost of such calculation is already very high with just 2 parameters. This serves as an additional benchmark and indeed confirms that the optimization procedure correctly converges to one of the maxima requiring much less simulations than the grid scan (for this example we performed a $9\times9$ parameter grid scan using 81 simulations while optimization required only 24 simulations). 

As already noted above, the investigation of the dependence of the discernible signal on the two independent probe waist sizes identified two particularly promising configurations. These are not equivalent because the studied collision scenario has two asymmetries: 1) Collision angles different from  $180^{\circ}$ affect the overlap volume of the pump and probe with an elliptic waist, and hence the signal photon amplitude (this effect becomes increasingly pronounced for collision angles closer to $\vartheta_{\text{col}}\to90^{\circ}$). Recall that the results discussed here are for $\vartheta_{\text{col}}=160^\circ$. 2) The orientation of the elliptic cross section of the probe relative to its polarization sets up a preferred direction. 

We separate the effects 1) and 2) by choosing $\vartheta_{\text{col}}=180^{\circ}$ and comparing the results with those for $\vartheta_{\text{col}}=160^{\circ}$.
Figure \ref{fig:theta_180_polarization} demonstrates this polarization dependence. In all considered cases the pump is polarized perpendicular to the probe. When orienting the long axis of the probe waist cross section along the probe polarization, we obtain a $\approx$ 2.3\% improvement in the total discernible signal compared to the situation where the probe polarization is oriented along the short ellipse axis. The difference becomes less pronounced when widening the short ellipse axis, i.e. for weaker focusing in this direction.
For $\vartheta_{\text{col}} = 160^{\circ}$ the polarization dependence is mixed with the collision angle asymmetry which dominates up to a certain waist size: for $\vartheta_{\text{col}} = 180^{\circ}$ the $\mu_x=1$ case always yields a smaller discernible signal than the one with $\mu_y=1$, but for $\vartheta_{\text{col}}=160^{\circ}$ this behavior is reversed up to $\mu_{x,y} \approx 12$.

\begin{figure}
    \centering
    \includegraphics[width=\linewidth]{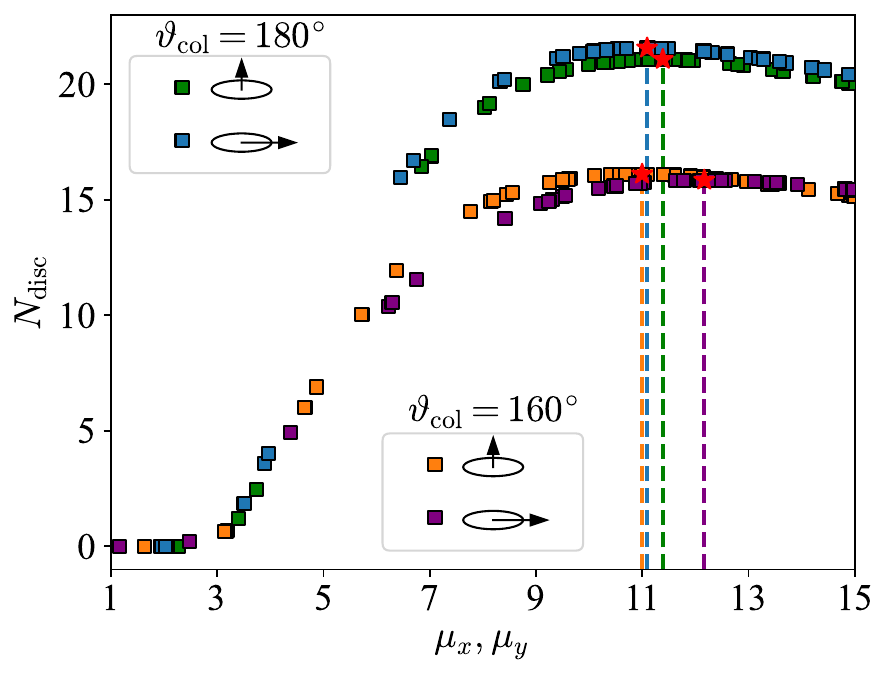}
    \caption{Number of discernible signal photons as a function of one elliptic waist size. The results presented here are for two beams of frequency $\omega_0$ colliding at an angle of $\vartheta_{\text{col}}=180^\circ$ (green, blue) and $\vartheta_{\text{col}}=160^\circ$ (orange, purple). Each beam has energy $W=25\,{\rm J}$ and duration $\tau=25\,{\rm fs}$. The probe beam has an elliptic cross section with either fixed $\mu_x=1$, and variable $\mu_y$, or variable $\mu_x$ and  fixed $\mu_y=1$. In all considered cases the probe is polarized perpendicular to the pump and oriented along the $x$ axis, which thus implies either an orientation along the short or long ellipse axis, respectively. The insets visualize the polarization direction relative to the ellipsis orientation for different datasets.} 
    \label{fig:theta_180_polarization}
\end{figure}

Finally, to demonstrate that optimization can deal with large parameter spaces, we consider the following scenario: two laser beams with variable elliptic cross sections collide under an angle of $\vartheta_{\text{col}} = 180^{\circ}$. Their total energy is fixed to $W_0=50\,{\rm J}$ but the energy distribution between the pulses is variable. This defines a 5-dimensional optimization problem where the free parameters are the two independent waist sizes of both beams parameterized by $\mu_{x,\ell}$, $\mu_{y,\ell}$ with $\ell\in\{1,2\}$ and the energy distribution between the beams. Here we artificially constrain the whole parameter space to $0\leq\{\mu_{x,l},\mu_{y,l}\}\leq9$ for convenience because the example is for demonstration purposes only; in general this is not a requirement.

\begin{figure*}
    \centering
    \includegraphics[width=\linewidth]{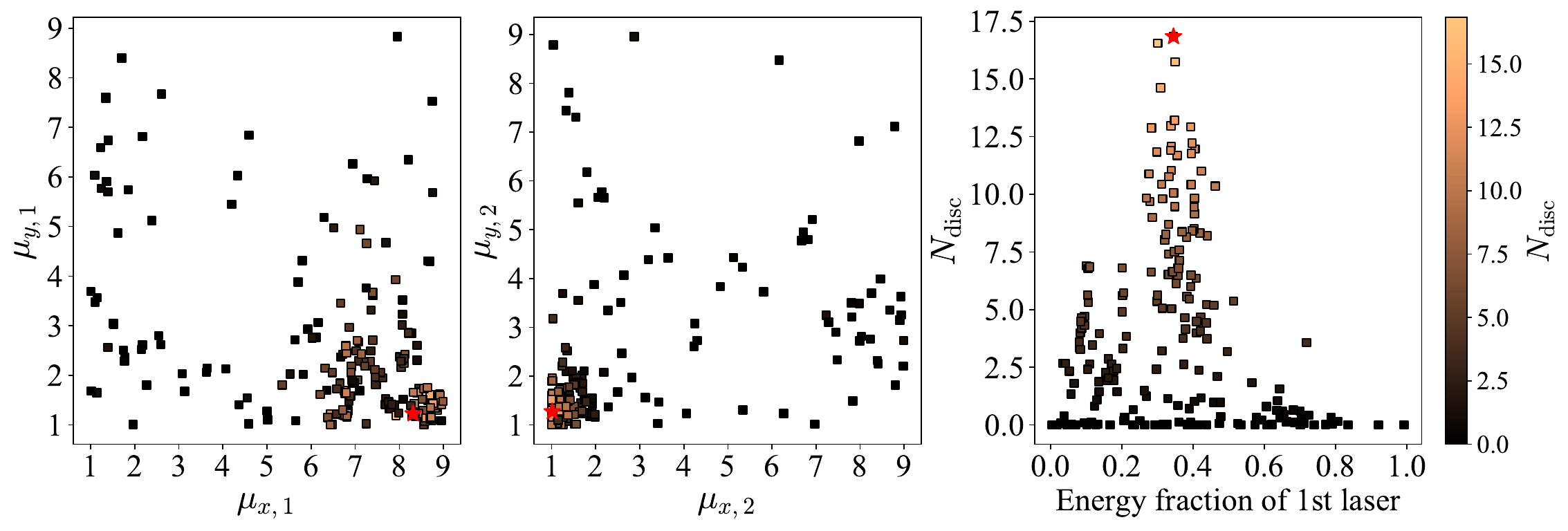}
    \caption{Optimization results in 5-dimensional parameter space: elliptic waist sizes of two beams ($\mu_{x,\ell}, \mu_{y,\ell}$) with $\ell \in \{1,2\}$ and energy distribution between the beams. Two beams (duration $\tau = 25$ fs each, their total energy is fixed to 50 J) of frequency $\omega_0$ collide at an angle of $\vartheta_{\text{col}}=180^\circ$.}
    \label{fig:5d_opt}
\end{figure*}

Figure \ref{fig:5d_opt} shows the results of optimization as projections of the constrained $d=5$ dimensional space to 2-d (the $\mu_{x,\ell}, \mu_{x,\ell}$ planes) or 1-d (energy fraction of the $\ell=1$ beam) parameter spaces.
Optimization suggests that one beam should have an elliptic cross section which is strongly focused along one axis and weakly focused along the perpendicular one (left plot).
At the same time, the other beam should be focused to its diffraction limit (middle plot) and should contain $2/3$ of initial energy budget (right plot). 
This results in the following optimal values for the five free parameters: $\mu_{x,1} \approx 8.31$, $\mu_{y,1} \approx 1.23$, $\mu_{x,2} \approx 1.27$, $\mu_{y,2} \approx 1.01$ and $W_{1} \approx 0.35\,W_0$.  

It is typically unfeasible to investigate such large parameter landscape with regular grid scans but physical intuition can give us guidance about the expectations for some of the free parameters. Namely, the scaling of the quantum vacuum signals in Eq.~\eqref{eq:amplitude} with powers of the field strength suggests that to maximize the signal one needs strong fields in the interaction volume. This requires at least one of the beams to be focused to its diffraction limit (this beam is then usually referred to as ``pump'' because it maximizes the signal photon amplitude that scales quadratically with the field of this beam and linearly with that of the other one). In the pump-probe collision scenario with the probe being the less focused pulse, the signal photon number in the probe channel emitted over all possible emission directions scales proportionally to $W_{\text{probe}} W_{\text{pump}}^2$. For fixed total energy $W_0$ this expression is maximized when $W_{\text{probe}} = \frac{1}{3} W_0, W_{\text{pump}} = \frac{2}{3} W_0$. We expect the optimal energy distribution to remain close to these values even for the discernible signal; cf. also \cite{karbstein2018vacuum}. At the same time, the elliptic waist sizes should agree with the results already discussed above. Overall, we might have expected the optimum to be at ($\mu_{x,1} = 9, \mu_{y,1} = 1, \mu_{x,2} = 1, \mu_{y,2} = 1, W_{1} = \frac{1}{3} W_0$). The optimization procedure did not converge to (but closely approached) this guess.

In summary, in this section we showed that self-consistent numerical simulations coupled to optimization provide more accurate and reliable estimates for the parameters maximizing the total discernible signal.

\section{Coherent harmonic focusing}
\label{sec:harmonics}

In the \textit{coherent harmonic focusing} (CHF) scenario considered here one of the two colliding laser pulses consists of $n$ harmonics ($\omega_0, 2\omega_0, ..., n\omega_0$) which are focused to their respective diffraction limits $w_{0,n}=\lambda_n=2\pi/(n\omega_0)$; cf. also \cite{gordienko2004relativistic, gordienko2005coherent}. Each pulse has a duration of $\tau = 20\,{\rm fs}$ while the energy distribution between the pulses is variable. Moreover, here we exclusively stick to a counter-propagating geometry with $\vartheta_{\text{col}} = 180^{\circ}$.

In the previous section we studied a scenario which allowed for a quantitatively accurate semi-analytic study of the discernible signal for collision angles sufficiently different from $180^{\circ}$ by using an infinite Rayleigh range approximation. Due to the various frequency components involved, for the coherent harmonic focusing studied below such treatment is more complicated. Following \cite{karbstein2019boosting} we envision to collide two optical pulses one of which undergoes coherent harmonic focusing before the collision. This partitions the pulse energy into different frequency modes and thereby generically enhances the peak field amplitude in the interaction volume and opens up additional inelastic signal channels giving rise to potential discernible signals. Since all pulses are focused to their diffraction limits and $\lambda_n<\lambda_0$ for $n>0$, particularly the elastic signal photon scattering channels associated with $n>1$ are expected to be not visible against the laser background because of their similar far-field divergences; cf. Eq.~(4) in the supplementary material of \cite{karbstein2019boosting}. However, a discernible signal at an energy $\approx 2\omega_0$ emitted at a polar angle $\vartheta \approx 90^{\circ}$ was identified. The microscopic origin of this channel was not analyzed in \cite{karbstein2019boosting}.

\begin{figure}[b!]
    \centering
    \includegraphics[width=\linewidth]{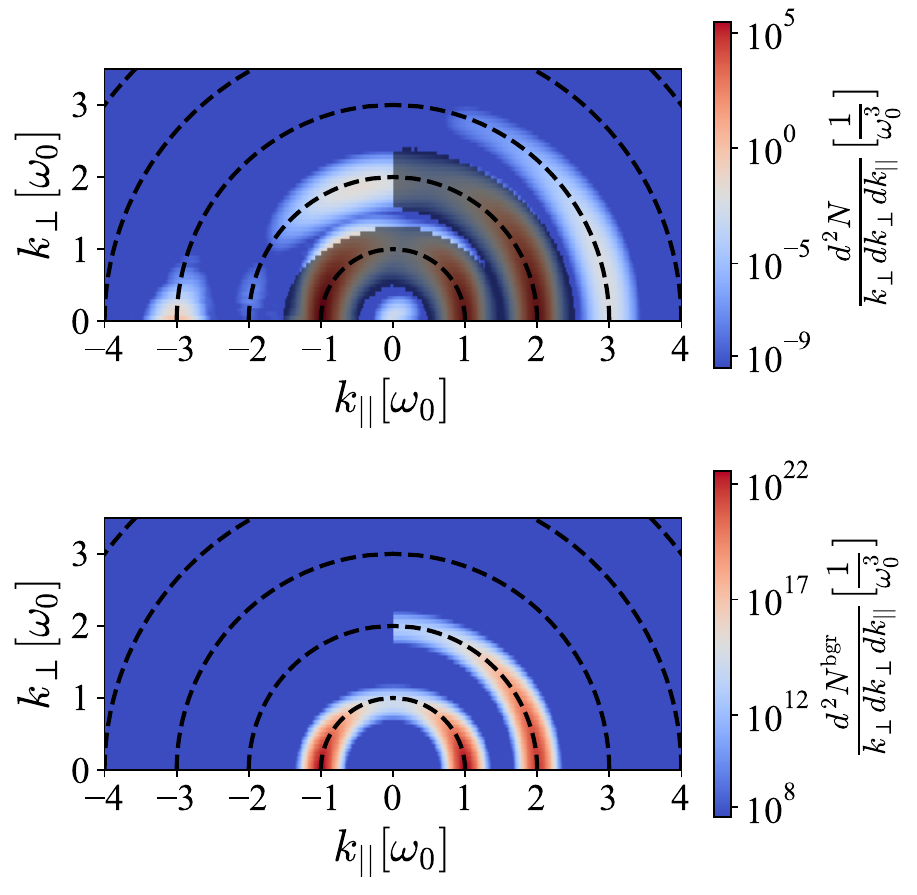}
    \caption{Differential distribution of signal (top) and background (bottom) photons integrated over the azimuthal angle ($k_{||} = k_z, k_{\perp} = \sqrt{\omega^2 - k_{||}^2} > 0$); the area where the signal is not discernible against the laser background is shaded. The CHF pulse consisting of two harmonics ($\omega_0, 2\omega_0$) propagates along the $z$-axis; each harmonic has an energy $W=25\,{\rm J}$, a duration $\tau=20\,{\rm fs}$ and is focused to its respective diffraction limit $w_{0,n}=\lambda_n$. The counter-propagating fundamental frequency $\omega_0$ pulse has the same parameters ($W=25\,{\rm J}$, $\tau=20\,{\rm fs}$, $w_{0}=\lambda_0$). The spectral laser pulse model \cite{waters2017beam,blinne2019all} was used to initialize the laser fields, which leads to the sharp cut-offs near $\vartheta = 90^{\circ}$ on the background plot.} 
    \label{fig:harmonics/cylindrical}
\end{figure}

To study the origin of this particular signal, we consider a simplified version of the coherent harmonic focusing scenario of \cite{karbstein2019boosting}: a pulse consisting of two harmonics ($\omega_0, 2\omega_0$) is propagating along the $z$-axis and collides with a counter-propagating fundamental frequency $\omega_0$ pulse. 
Figure~\ref{fig:harmonics/cylindrical} shows the differential distribution of signal and background photons in this scenario. Because the discernible $2\omega_0$ signal appears at $\vartheta>\sqrt{3}\,\theta$, with radial divergence $\theta\simeq1/\pi$ of the coherent harmonic beam, it should originate from a manifestly inelastic scattering process characterized by frequency transfers from both colliding laser fields. A promising candidate to arrive at such a signal is the channel $\omega=\omega_0+2\omega_0-\omega_0$ which describes the merging of a laser photon of the fundamental frequency pulse with one from the second harmonic of the CHF pulse accompanied by an emission into the fundamental frequency mode of the CHF pulse. In \cite{karbstein2019boosting} the authors studied this scenario with a CHF pulse comprising up to 12 harmonics. In this case, each new harmonic provides additional ways to contribute to the discernible $\omega\approx2\omega_0$, $\vartheta\approx90^\circ$ signal via the above channel, because $n\omega_0+\omega_0-(n-1)\omega_0=2\omega_0$, etc.    

\begin{figure}[]
    \centering
    \includegraphics[width=\linewidth]{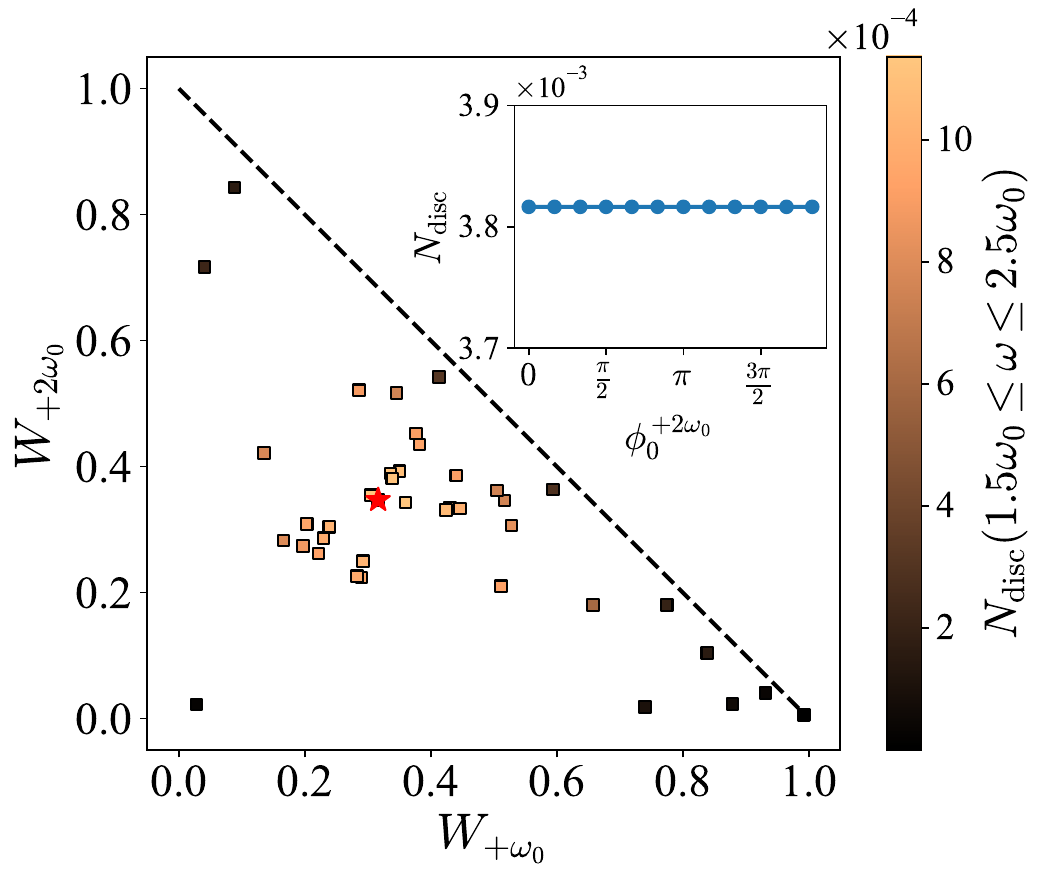}
    \caption{Optimization of energy distribution in three frequency modes: a CHF pulse consisting of two harmonics (energies $W_{+\omega_0}$ and $W_{+2\omega_0}$) and a counter-propagating pulse (energy $W_{-\omega_0}$). The total energy is fixed to $W_0=50\,{\rm J}$, both pulses have a duration of $\tau=20\,{\rm fs}$ and each harmonic is focused to its diffraction limit. The discernible signal was calculated only for energies $1.5\, \omega_0 \leq \omega \leq 2.5 \,\omega_0$ to capture the channel of interest. Optimization suggests that to maximize the signal in the $\omega \approx\omega_0$, $\vartheta\approx90^\circ$ channel one needs to distribute the energy equally between all three pulses: $(W_{+\omega_0}, W_{+2\omega_0}, W_{-\omega_0}) \approx (1/3, 1/3, 1/3)\, W_0$. The inset shows the total discernible signal in this channel as a function of phase delay of the $+2\omega_0$ frequency component.}
    \label{fig:harmonics/21-1_W_opt}
\end{figure}

Optimization methods can help us to test our hypothesis and understand where this signal comes from. Fixing the total laser pulse energy to $W_0=50\,{\rm J}$, we describe the energy distribution between the two modes of the CHF pulse and the fundamental frequency pulse with two free parameters: the energy put into the fundamental mode of the CHF pulse $W_{+\omega_0}$ and the energy of its 2nd harmonic $W_{+2\omega_0}$ (the energy of the fundamental frequency pulse is then given by $W_{-\omega_0} = W_0 - W_{+\omega_0} - W_{+2\omega_0}$). Information about the particular channel giving rise to the discernible signal can be encoded in its dependency on the energy of the driving laser pulses. If the three laser photons forming this signal come from different modes then the total signal in this channel should depend linearly on the energy of each mode and its maximum should be found at $W_{+\omega_0}=W_{+2\omega_0}=W_{-\omega_0} = W_0/3$. This is illustrated by our optimization results in Fig. \ref{fig:harmonics/21-1_W_opt} where the signal in the specified channel is indeed maximized for an equal energy distribution between the three different frequency modes. 

Whenever a signal originates from several channels, their contributions interfere. In turn, changing the phase between different contributions would affect the signal. As an additional check, the inset in Figure \ref{fig:harmonics/21-1_W_opt} shows the dependence of the discernible signal at $\omega\approx2\omega_0$, $\vartheta\approx90^\circ$ as a function of the phase delay of the $+2\omega_0$ contribution. This confirms that the signal is not affected by the phase delay and thus supports our above hypothesis.

In this section optimization clearly helped us to identify the microscopic origin of the discernible signal.

\section{Conclusion}
\label{sec:conclusion}
The development and consistent analysis of modern theoretical and experimental setups in the research area of light-by-light scattering generically requires more accurate simulations of collision scenarios and optimization over large parametric spaces.
In particular, the {\it discernible} quantum vacuum signal which constitutes the quantity that is most relevant for experiment depends non-trivially on the collision geometry and parameters of the driving laser fields.
To quantitatively explore these large parameter spaces, human physical intuition is not enough and standard grid scans require too much computational resources. Therefore, ultimately some efficient automation is required both for exploratory searches and finding the best experimental parameters in a fixed collision setting. 

To exemplify this, in the present article we used modern optimization methods (Bayesian optimization) for different light-by-light scattering scenarios.
Resorting to known results for a specific scenario previously studied in the literature \cite{gies2022all}, we verified that the outcome of optimization agrees with previous findings obtained by naive grid scan simulations.
In this context, we identified the true optimal choice for the elliptic focus cross section of the probe in a two-beam collision scenario. This choice yields a total discernible signal that is essentially twice as large as the previously predicted maximum.
We explicitly demonstrated that the use of optimization to constrain relevant parameter spaces is much more efficient than conventional grid scans. Optimization allows to study higher dimensional problems and find optimums with greater accuracy.

Aside from that, we also studied the coherent harmonic focusing scheme suggested by \cite{karbstein2019boosting} to boost quantum vacuum signals. Here, our main goal was to unveil the unknown nature of the microscopic channel giving rise to the discernible quantum vacuum signal in this setup; its origin was found with the help of optimization. In exploratory searches many different hypotheses should be tested and optimization can alleviate the computational burden by doing this very efficiently.    

Overall, we firmly believe that optimization methods will prove to be really useful in particular for future numerical studies and towards the design of dedicated light-by-light scattering experiments at high-field facilities. At the same time, its huge potential is certainly not limited to that and will also assist the study of many other strong fields QED phenomena. 

\acknowledgements

We developed and used the package \cite{github} to obtain the results for this article. 
The authors acknowledge the usage of the HPC cluster “DRACO” of the University of Jena for obtaining the numerical results presented in this article.
This work has been funded by the Deutsche Forschungsgemeinschaft (DFG) under Grant Nos. 416607684 and  416702141 within the Research Unit FOR2783/2.
The research leading to the presented results received additional funding from the European Regional Development Fund and the State of Thuringia (Contract No. 2019 FGI 0013).

%

\end{document}